\documentclass{webofc}
\usepackage[varg]{txfonts}   
\usepackage[utf8]{inputenc}
\begin{document}
\title{Constraining hydrostatic mass bias and cosmological parameters with the gas mass fraction in galaxy clusters}
%
%

\author{\firstname{Raphaël} \lastname{Wicker}\inst{1}\fnsep\thanks{\email{raphael.wicker@ias.u-psud.fr}} \and 
	\firstname{Marian} \lastname{Douspis}\inst{1} \and
	\firstname{Laura} \lastname{Salvati}\inst{1}
	\and
	\firstname{Nabila} \lastname{Aghanim}\inst{1}
}

\institute{Université Paris-Saclay, CNRS, Institut d’Astrophysique Spatiale, 91405, Orsay, France
}

\abstract{%
The gas mass fraction in galaxy clusters is a convenient tool to use in the context of cosmological studies.
Indeed this quantity allows to constrain the universal baryon fraction $\Omega_b/\Omega_m$, as well as other parameters like the matter density $\Omega_m$, the Hubble parameter $h$ or the Equation of State of Dark Energy $w$.

This gas mass fraction is also sensitive to baryonic effects that need to be taken into account, and that translate into nuisance parameters. 
Two of them are the depletion factor $\Upsilon$, and the hydrostatic mass bias $B = (1 - b)$. 
The first one describes how baryons are depleted in clusters with respect to the universal baryon fraction, while the other encodes the bias coming from the fact that the mass is deduced from X-ray observations under the hypothesis of hydrostatic equilibrium.

We will show preliminary results, obtained using the {\it Planck}-ESZ clusters observed by XMM-{\it Newton}, on both cosmological and cluster parameters.
We will notably discuss the investigation on a possible redshift dependence of the mass bias, which is considered to be non-existent in hydrodynamic simulations based on $\Lambda$-CDM, and compare our results with other studies.

}
\maketitle
\section{Introduction}
\label{intro}
Being the most massive gravitationally bound systems of our universe, galaxy clusters carry a lot of information. 
They can notably be used as powerful cosmological probes \cite{1993Natur.366..429W}, or as astrophysical objects of study, to better understand the physics of the intra-cluster medium.
Their baryonic component being mainly under the form of hot gas \cite{2004ApJ...617..879L}, of which the fraction is assumed to be relatively well known and understood, the gas mass fraction of galaxy clusters can be used as a robust cosmological probe.
Indeed this gas fraction is considered to be a good proxy for the universal baryon fraction \cite{2011ASL.....4..204B}.

Clusters being the siege of astrophysical phenomena, the gas mass fraction is also sensitive to the baryonic physics inside these objects.
Baryonic physics are encoded for one part in the depletion factor $\Upsilon$, which describes how the gas is depleted with respect to the universal baryon fraction \cite{2008MNRAS.383..879A}, and for the other part in the hydrostatic mass bias $B = (1-b) = \frac{M_{measured}}{M_{true}}$.
This bias comes into play when measuring the cluster masses from X-ray or SZ observations. 
Indeed such measures assume that clusters are at the hydrostatic equilibrium, which is not entirely true due to a fraction of non-thermal pressure support inside these objects.
Several works have been studying the hydrostatic mass bias (see e.g. \cite{2019A&A...626A..27S}, \cite{2019SSRv..215...25P}), and notably its evolution with the redshift.
Our goal in this work is to try and compare the findings of these studies with results obtained using the gas mass fraction in clusters. 
\section{Data}
\label{sec-1}
We use gas masses and total masses at $R_{500}$ from 120 clusters of the {\it Planck}-ESZ sample \cite{2011A&A...536A...8P} as seen in follow-up XMM-{\it Newton} observations analysed in \cite{2020ApJ...892..102L}. These clusters span the redshift range $[0.059; 0.546]$, with their total mass derived from the hydrostatic equilibrium equation given below :
\begin{equation}
    M_{tot} (<r) = - \frac{r k_B T(r)}{G \mu m_p} \left (\frac{d \ln{\rho(r)}}{d \ln{r}}  + \frac{d \ln{T(r)}}{d \ln{r}}\right)
    \label{hydro_mass}
\end{equation}
From the gas masses and total masses, we can compute the gas fraction of these clusters, $f_{gas} = M_{gas}/M_{tot}$,  displayed in Figure 1 below.
\label{eq-1}
\begin{figure}[h!]
\centering
\sidecaption
\includegraphics[trim = {1cm 6cm 1cm 1.5cm}, width = 0.7\textwidth, height = 5cm]{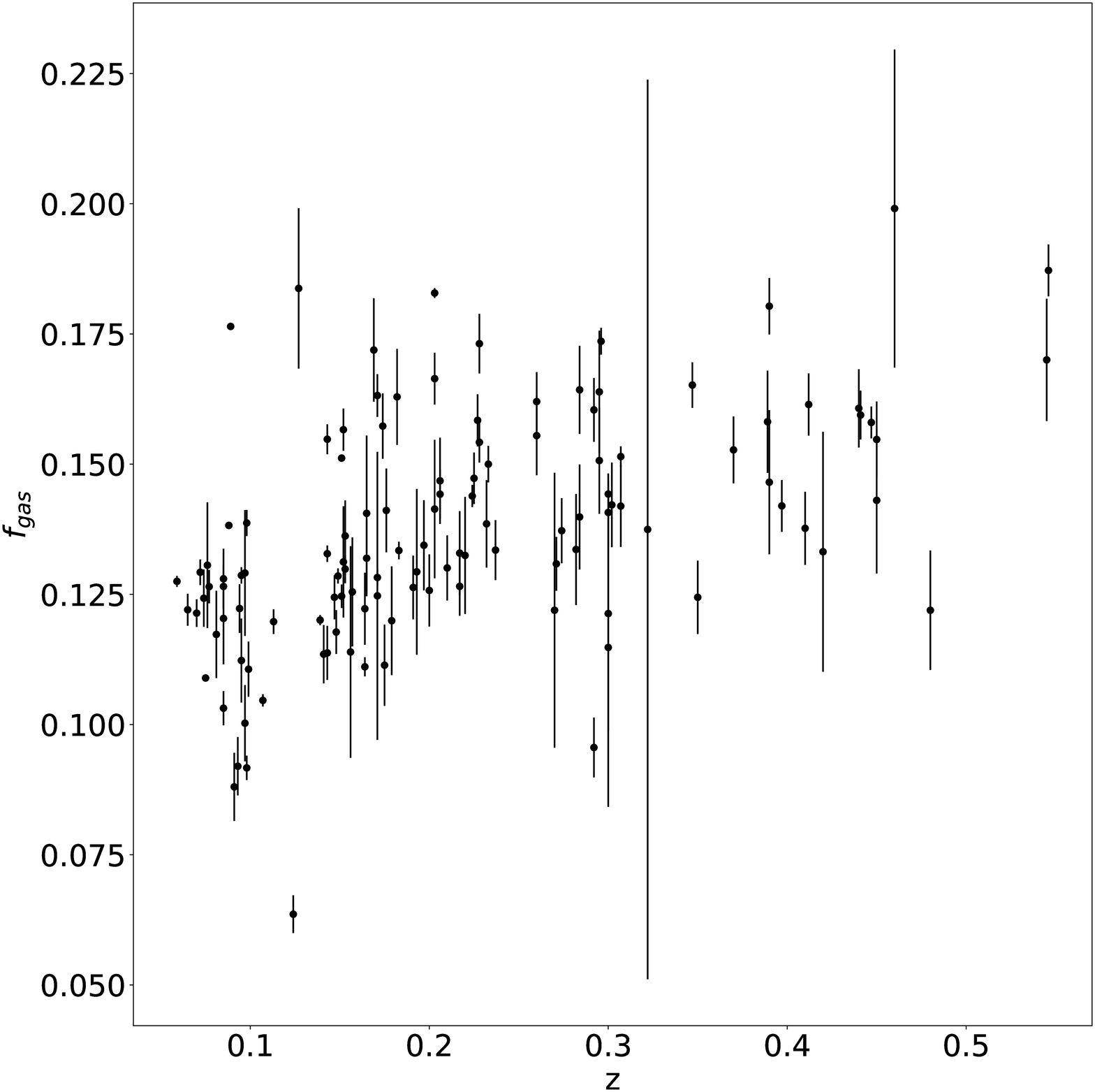}
\label{sample}
\caption{Observed gas fraction of the {\it Planck}-ESZ sample}
\end{figure}
\section{Modelling the redshift evolution of the bias}
\label{sec-2}
The redshift evolution of the gas mass fraction depends on several instrumental, astrophysical, and cosmological contributions, which can be summed up in equation \ref{fgas_z} below, from \cite{2008MNRAS.383..879A}:
\begin{equation}
    f_{gas}(z) = K \times \frac{\Upsilon(z)}{B(z)} \times A(z) \times \left ( \frac{\Omega_b}{\Omega_m}\right) \times \left ( \frac{D_A^{fid}(z)}{D_A(z)}\right)^{3/2}
    \label{fgas_z}
\end{equation}
where $K$ is an instrumental calibration constant (assumed to be 1 in this work), $\Upsilon$ is the baryon depletion factor, and $B$ is the hydrostatic mass bias $B = (1-b)$. $A(z)$ is an angular correction parameter, constant and equal to 1 if we are in the fiducial cosmology considered. $(\Omega_b/\Omega_m)$ is the universal baryon fraction, and $D_A$ is the angular diameter distance. 
In this analysis we consider as fiducial cosmology a flat $\Lambda$-CDM with $h = 0.7$, $\Omega_m = 0.3$ and $\Omega_\Lambda = 0.7$.

As shown in equation \ref{fgas_z}, what we actually constrain when looking at the redshift evolution of $f_{gas}$ is the evolution of the ratio $\Upsilon(z)/B(z)$. 
Based on hydrodynamical simulations from \cite{2013MNRAS.431.1487P}, we assume a constant depletion factor $\Upsilon(z) = \Upsilon_0$. For a study of the evolution of $\Upsilon(z)$ in galaxy cluster data, see \cite{2021EPJC...81..296B}. 
We therefore assume that all the evolution comes from the bias, for which we assume a linear evolution with a constant term $B_0$ (the bias at $z = z_{pivot}$) and a slope $B_1$ :
\begin{equation}
    B(z) = B_0 + B_1 \times (z - z_{pivot})
\end{equation}
with $z_{pivot} = \left<z\right> = 0.218$.
In a first part of the study we consider a cosmology fixed at the values of \cite{2020A&A...641A...6P}.
We thus fit the measured $f_{gas}$ with a MCMC using the python package {\tt emcee}, with the set of priors given in Table \ref{priors-1st-part} and adding a term accounting for the intrinsic scatter in the data, $\sigma_f$.
\begin{table}
\centering
\caption{Set of priors used in the first part of the analysis. $\mathcal{U}(l,u)$ means a uniform prior of lower bound $l$ and upper bound $u$, while $\mathcal{N}(\mu, \sigma)$ means a normal prior of mean $\mu$ and standard deviation $\sigma$. The prior on $\Upsilon_0$ comes from \cite{2013MNRAS.431.1487P}.}
\label{priors-1st-part}       
\begin{tabular}{c|c}
\hline
$B_0$ & $\mathcal{U}(0.3, 1.7)$ \\
$B_1$ & $\mathcal{U}(-1.5, 1.5)$ \\
$\Upsilon_0$ & $\mathcal{N}(0.85, 0.03)$ \\
$\sigma_f$ & $\mathcal{U}(0, 1)$\\\hline
\end{tabular}
\end{table}
\section{Results}
\label{sec-4}
Matching the selection from weak lensing studies like the {\sc CoMaLit} \cite{2017MNRAS.468.3322S} or {\sc LoCuSS} \cite{2016MNRAS.456L..74S} studies, we chose to study the subsample $z > 0.2$, as well as our full sample.
This choice was also motivated by the results from \cite{2019A&A...626A..27S} showing that the trends in the mass bias depended on the considered redshift range, with results changing when considering only clusters with $z>0.2$.
\subsection{Full sample}
\label{subsec-4-1}
We first carried out our analysis on the full sample, with redshifts in the range [0.059, 0.546]. We kept all the cosmological parameters fixed at their values from \cite{2020A&A...641A...6P}, leaving free the constant term of the bias $B_0$, the slope of the bias $B_1$, and the intrinsic scatter $\sigma_f$. The depletion factor $\Upsilon$ is treated as a nuisance parameter.
Using this model, we obtained the plot given in Figure 2. 
\begin{figure}[h!]
\label{full_sample_contours}
\centering
\sidecaption
\includegraphics[width = 0.45\textwidth, height = 5cm, trim = {0 1cm 1cm 0.5cm}]{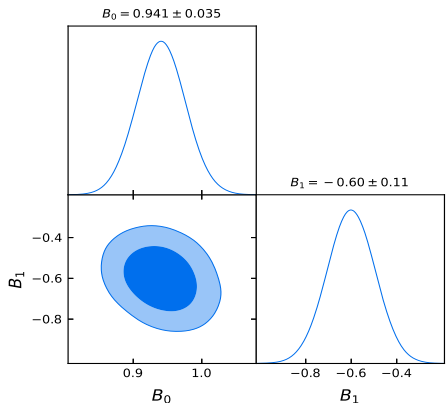}
\caption{Constraints on $B_1$ for the full sample. Shaded areas mark 1 and 2$\sigma$ regions while the numeric values give the uncertainties at $1\sigma$.}
\end{figure}
In particular with $B_1 = -0.60 \pm 0.11$ we obtain a slope very different from 0, hinting at a redshift evolution of the mass bias at 5.5$\sigma$, with a bias increasing with the redshift. Indeed with $B_1 < 0$, we have a parameter $B(z) = (1-b)$ decreasing with the redshift, meaning that the masses are more biased at higher redshift. These trends are in agreement with \cite{2017MNRAS.468.3322S} and \cite{2016MNRAS.456L..74S} based on weak lensing studies, but in contradiction with \cite{2019A&A...626A..27S} based on tSZ number counts with almost the same clusters as this work.
\subsection{High redshift clusters}
\label{subsec-4-2}
Following \cite{2019A&A...626A..27S} we now focus on the clusters for which z > 0.2, using the same model as in Sect. \ref{subsec-4-1}. We obtain the contours from Figure 3.
\begin{figure}[h!]
\label{highz_sample_contours}
\centering
\sidecaption
\includegraphics[width = 0.45\textwidth, height = 5cm, trim = {0 1cm 1cm 1cm}]{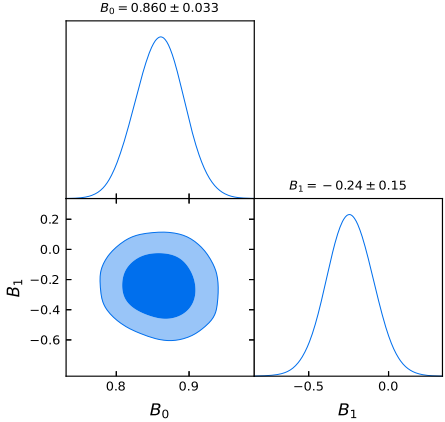}
\caption{Constraints on $B_1$ for clusters with $z>0.2$. Shaded areas mark 1 and 2$\sigma$ regions.}
\end{figure}

We find this time $B_1 = -0.24 \pm 0.15$ which means a slope much closer to 0, with this time a mild $1.7\sigma$ discrepancy with no evolution of the bias. 
This behaviour had already been noted in \cite{2019A&A...626A..27S}, which found no evolution of the bias within $1\sigma$ when focusing only on the clusters with z > 0.2.
In addition, the value of $B_0 = 0.860 \pm 0.035$ is also fully compatible with other values of $B = (1-b)$ from weak-lensing studies, like \cite{2020MNRAS.497.4684H} or obtained from hydrodynamical simulations, like \cite{2016ApJ...827..112B}.
\section{Impact on cosmological analysis}
\label{sec-5}
In a second part of the analysis we let some of the cosmological parameters free. 
In this preliminary work we focused only on the high redshift clusters. A more complete study focused on different subsamples and selections in mass and redshift are proposed in Wicker et al. {\it in prep}.
We first considered no evolution of the hydrostatic mass bias, then applied a non-zero $B_1$, based on the values derived in the previous part of this work. 
The set of priors considered for this work is the following:
\begin{table}[h!]
\centering
\caption{Set of priors used in the second part of the analysis. $\mathcal{U}(l,u)$ means a uniform prior of lower bound $l$ and upper bound $u$, while $\mathcal{N}(\mu, \sigma)$ means a normal prior of mean $\mu$ and standard deviation $\sigma$. The prior on $B_0$ comes from \cite{2015MNRAS.449..685H}, and the prior on $h$ comes from \cite{2020A&A...641A...6P}. The priors on $\Upsilon_0$ and $\sigma_f$ are the same as in Table 1.}
\label{priors-2nd-part}       
\begin{tabular}{c \space c}
\hline \hline
$B_0$ & $\mathcal{N}(0.780, 0.092)$\\
$B_1$ & $\mathcal{N}(0, 0.001)$ (Sect. \ref{subsec-5-1}) then $\mathcal{N}(-0.24, 0.15)$ (Sect. \ref{subsec-5-2})\\
$\Omega_b/\Omega_m$ & $\mathcal{U}(0.05, 0.3)$ \\
$\Omega_m$ & $\mathcal{U}(0.01, 1.0)$ \\
$h$ & $\mathcal{N}(0.674, 0.005)$ \\
\hline
\end{tabular}
\end{table}
\subsection{Considering no evolution of the bias}
\label{subsec-5-1} 
We first consider the hydrostatic bias to be constant, and set a gaussian prior centered on the value from the weak lensing study CCCP \cite{2015MNRAS.449..685H}.  We let the matter density $\Omega_m$ and the universal baryon fraction $\Omega_b/\Omega_m$ free, while the Hubble parameter $h$ is set at its value from \cite{2020A&A...641A...6P}. 

The output of our MCMC on this model is shown in Figure 4 below. The values from {\it Planck} 2018 cosmological parameters \cite{2020A&A...641A...6P} are represented as the orange bands for $\Omega_m$ and $\Omega_b/\Omega_m$.
\begin{figure}[h!]
\label{constantB_cosmocontours}
\centering
\sidecaption
\includegraphics[width = 0.55\textwidth, height = 7cm, trim = {0 1cm 1.5cm 1cm}]{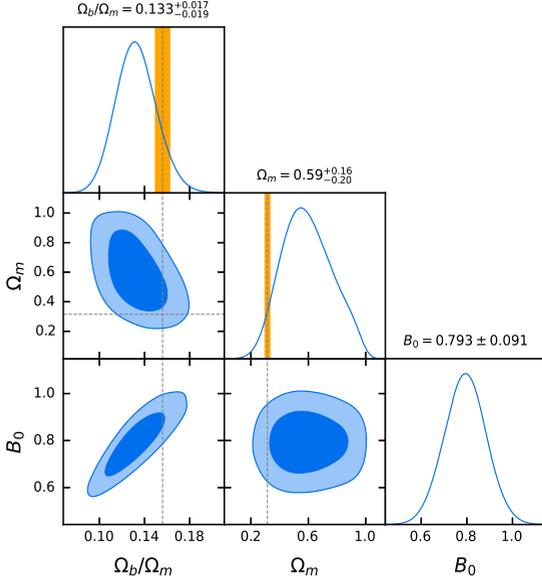}
\caption{Contours at 1 and 2$\sigma$ and marginalized posteriors for the cosmological study when assuming no redshift evolution of the bias.}
\end{figure}
We show that when we consider no evolution of the bias, the derived value of $\Omega_b/\Omega_m = 0.133^{+0.017}_{-0.019}$ is compatible with the value from {\it Planck}, yet peaks lower, and we can see a degeneracy between $B_0$ and $\Omega_b/\Omega_m$, as expected from equation \ref{fgas_z}. On the other hand we can see that the posterior for $\Omega_m$ is very wide and peaks high above the {\it Planck} value, as we obtain $\Omega_m = 0.59^{+0.16}_{-0.20}$.
\subsection{Considering an evolution of the bias}
\label{subsec-5-2} 
When we consider an evolution of the bias with the redshift, keeping the same prior on $B_0$, we get back to values that are much more compatible with the {\it Planck} values, as we can see in Figure 5. Indeed we find $\Omega_b/\Omega_m = 0.144^{+0.018}_{-0.021}$ and $\Omega_m = 0.35^{+0.17}_{-0.25}$, although the latter peaks slightly below the {\it Planck} value of 0.315 \cite{2020A&A...641A...6P}. 
We can also see that just like $B_0$ and $\Omega_b/\Omega_m$, $\Omega_m$ and $B_1$ are strongly degenerated, explaining the strong impact on the constraints of this parameter when we go from a constant bias to a varying bias.
\begin{figure}[h!]
\label{varyingB_cosmocontours}
\centering
\sidecaption
\includegraphics[width = 0.65\textwidth, height = 8cm, trim = {0 2cm 2cm 0cm}]{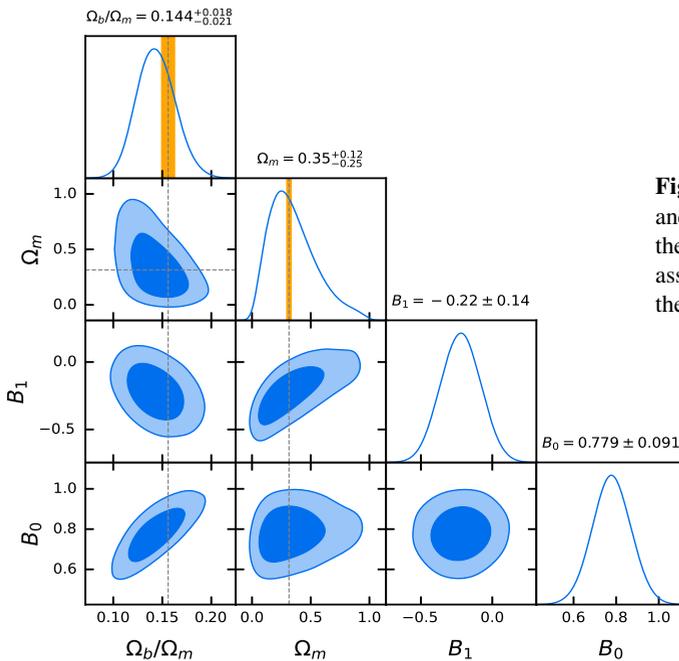}
\caption{Contours at 1 and 2$\sigma$ and marginalized posteriors for the cosmological study when assuming a redshift evolution of the bias.}
\end{figure}
\section{Conclusion}
As we have shown, our results on the redshift evolution of the mass bias are strongly dependent on the redshift range at which we chose to study our clusters, hinting at a strong sample dependence.
To try and mitigate this effect, a similar work could be carried out on a larger sample. 
We would also need to take into account possible mass selection effects, as cluster at higher redshifts in the sample are also of higher mass, and the hydrostatic mass bias may also be mass dependent \cite{2019A&A...626A..27S}.
The paper from Wicker et al. {\it in prep} will propose a complete study of the mass and redshift evolution of fgas in galaxy clusters, and as such of the hydrostatic mass bias, based on the method described in these proceedings.
Finally, we have shown that these results on the evolution of $B$ with redshift absolutely need to be taken into account when carrying out a cosmological analysis, as disregarding such evolution could strongly bias the quality of the constraints obtained in the end.
As it turns out, most cosmological studies using $f_{gas}$ are carried out at $R_{2500}$ instead of $R_{500}$. 
Using high spatial resolution measurements in X-rays {\it and} in SZ, which could be allowed by NIKA2, could help to constrain better the evolution of the mass bias at these radii.







\end{document}